\documentclass{MicMalChal}

\usepackage[normalem]{ulem}
\usepackage{algorithm}
\usepackage{algpseudocode}
\usepackage{pifont}
\usepackage[table]{xcolor}
\usepackage{multirow}
\usepackage{graphicx}
\usepackage{tikz}
\usepackage{pgfplots}
\usepackage{pgfplotstable}
\usepackage{comment}
\pgfplotsset{compat=newest}
\usepackage{caption}
\usepackage{tcolorbox}
\usepackage{subcaption}
\usepackage{url}
\usepackage{breakurl}

\newcommand{\MA}[1]{\textbf{ \textcolor{green}{\\MA: #1\\}}}

\usepackage{tabto}

\begin{document}
%

\conferenceinfo{CODASPY}{'16, March 09-11, 2016, New Orleans, LA, USA}
\CopyrightYear{2016} 

\title{Novel Feature Extraction, Selection and Fusion for Effective Malware Family Classification}

\numberofauthors{5} 
%
\author{
\alignauthor
Mansour Ahmadi\\
       \affaddr{Department of Electrical and Electronic Engineering}\\
       \affaddr{University of Cagliari, Italy}\\
       \email{mansour.ahmadi@diee.\\unica.it}
\alignauthor
Dmitry Ulyanov\\
       \affaddr{Skolkovo Institute of Science and Technology, Russia}\\
       \email{dmitry.ulyanov@\\skolkovotech.ru}
\alignauthor
Stanislav Semenov\\
       \affaddr{National Research University Higher School of Economics, Russia}\\
       \email{stasg7@gmail.com}
\and
\alignauthor
Mikhail Trofimov\\
       \affaddr{Moscow Institute of Physics and Technology, Russia}\\
       \email{mikhail.trofimov@phystech.\\edu}
\alignauthor
Giorgio Giacinto\\
       \affaddr{Department of Electrical and Electronic Engineering}\\
       \affaddr{University of Cagliari, Italy}\\
       \email{giacinto@diee.unica.it}
}
\maketitle
\begin{abstract}
Modern malware is designed with mutation characteristics, namely polymorphism and metamorphism, which causes an enormous growth in the number of variants of malware samples. Categorization of malware samples on the basis of their behaviors is essential for the computer security community, because they receive huge number of malware everyday, and the signature extraction process is usually based on malicious parts characterizing malware families. Microsoft released a malware classification challenge in 2015 with a huge dataset of near 0.5 terabytes of data, containing more than 20K malware samples. The analysis of this dataset inspired the development of a novel paradigm that is effective in categorizing malware variants into their actual family groups. This paradigm is presented and discussed in the present paper, where emphasis has been given to the phases related to the extraction, and selection of a set of novel features for the effective representation of malware samples. Features can be grouped according to different characteristics of malware behavior, and their fusion is performed according to a per-class weighting paradigm. The proposed method achieved a very high accuracy ($\approx$ 0.998) on the Microsoft Malware Challenge dataset.
\\
\end{abstract}
\section{Introduction}
In recent years, malware coders developed sophisticated techniques to elude traditional as well as modern malware protection mechanisms. On the other hand, developers of anti-malware solutions need to develop counter mechanisms for detecting and deactivating them, playing a cat-and-mouse game. The huge number of malware families, and malware variants inside the families, causes a major problem for anti-malware products. For example, McAfee Lab's antimalware solutions reported more than 350M total unique malware samples in Q4 of 2014, that represents a growth of 17\% with respect to the analogous data in Q3 \cite{Mcafee:2015}. Symantec reported more than 44.5 million new pieces of malware created in May 2015 \cite{SymantecVariant}. Analyzing the malicious intent in this vast amount of data requires a huge effort by anti malware companies. One of the main reasons for this high volume of malware samples is the extensive use of polymorphic and metamorphic techniques by malware developers, which means that malicious executable files belonging to the same malware \texttt{family} are constantly modified and/or obfuscated. In particular, polymorphic malware has a static mutation engine that encrypts and decrypts the code, while metamorphic malware automatically modify the code each time it is propagated.
\\
Malware detection and classification techniques are two separate tasks, that are performed by anti-malware companies. Firstly, an executable needs to be analyzed to detect if it exhibits any malicious content: Then, in the case a malware is detected, it is assigned to the most appropriate malware family through a classification mechanism. There are various ways for detecting malware in the wild, and detecting a zero-day malware is still a challenging task. For example, Kaspersky recently discovered a new variant of Duqu, Duqu 2.0, in their own internal networks in July of 2015 \cite{Duqu2}. The detection of this kind of advanced malware is usually carried out within a sandbox environment by leveraging on a powerful heuristic engine. After the malware detection step, malware need to be categorized into groups, corresponding to their families, for further analysis. As far as a very high number of malware variants is concerned, the need for the automation of this process is clear-cut.
\\
The analysis of malicious programs is usually carried out by static techniques \cite{Sami:2010:MDB,Ye:2008:IntMal, Narouei:2015:DLLMiner} and dynamic techniques \cite{Willems:2007:CWSandbox, Rieck:2008:LCM, Ahmadi:2013:MDSP, Wuchner:2014:MDQ}. Analyzers extract various characteristics from the programs' syntax and semantic such as operation codes \cite{Santos:2013:opcode} and function call graph \cite{Hu:2009:LMI} from the disassembled code, or string signatures \cite{Griffin:2009:AGS} and byte code n-grams \cite{Tabish:2009:MDU, Nissim:2014:NovelActive} from the hex code, or different structural characteristics from the PE header, such as dependencies between APIs \cite{Ye:2008:IntMal} and DLLs \cite{Narouei:2015:DLLMiner}. Some other works \cite{Shafiq:2009:PE-Miner} also explored the analysis of metadata such as the number of bitmaps, the size of import and export address tables besides the PE header's content. The aforementioned content-based detection systems, like those considering bytecode n-grams, APIs, and assembly instructions, are inherently susceptible to false detection due to the fact of polymorphism and metamorphism. In addition, these techniques are not appropriated in the case of malware samples such as the one with \texttt{00yCuplj2VTc9ShXZDvnxz} hash name, that does not contain any APIs, and also contains a few assembly instructions because of packing.
\\
In this paper, we propose a learning-based system which uses different malware characteristics to effectively assign malware samples to their corresponding families without doing any deobfuscation and unpacking process. Although unpacking may lead to the extraction of more valuable features if the packers are known, unpacking is a costly task, and dealing with customized packers is even more challenging. Hence, we aim to perform classification without the need to unpack the sample. In addition, the system doesn't need to be evaluated on any packed goodware, because the problem of malware classification already assumes all of the samples to be malware. Finally, as this paper focuses on malware classification, we didn't make any analysis of evasion mechanisms employed to evade detection. 

For each malware sample, we compute not only a set of content-based features by relying on state-of-the-art mechanisms, but also we propose the extraction of powerful complementary statistical features that reflects the structure of portable executable (PE) files. The decision of not using more complex models like n-grams, sequences, bags or graphs, allowed us to devise a simple, yet effective, and efficient malware classification system. Moreover, we implemented an algorithm, inspired by the forward stepwise feature selection algorithm \cite{James:2014:ISL}, to combine the most relevant feature categories to feed the classifier, and show the trade-off between the number of features and accuracy. To better exploit both the richness of the available information, in the number of the malware samples for training the classifier, and the number of features used to represent the samples, we resorted to ensemble techniques such as bagging \cite{Kuncheva:2014:EM}. 

We evaluated our system on the data provided by Microsoft for their malware Challenge hosted at Kaggle\footnote{\texttt{https://www.kaggle.com/c/malware-classification}}, and achieved 99.77\% accuracy. The source code of our method is available online\footnote{\texttt{https://github.com/ManSoSec/Microsoft-Malware-Challenge}}.
\\
In summary, the original contributions of this paper are the following:
\begin{itemize}
\item The extraction and evaluation of different features based on the content and the structure of a malware that is performed directly on the packed executable file, and doesn't require the costly task of unpacking,
\item A novel technique that extracts information on the structural characteristics of PEs, to accurately classify even obfuscated malware,
\item The use of a limited number of features compared to other state-of-the-art systems, so that the method is apt to be used in large-scale malware categorization tasks,
\item An algorithm for feature fusion that outputs the most effective concatenation of features categories, each category being related to different aspects of the malware, thus avoiding the combination of all the possible feature categories, and providing a trade-off between accuracy and the number of features,
\item The assessment of the performances of the proposed malware classification on a dataset recently released by Microsoft, that can be considered one of the most updated and reliable testbeds for the task at hand.
\end{itemize}

The rest of the paper is organized as follows: a survey on the related work is presented in section 2; section 3 presents the details of the proposed method. Results of the experiments are discussed in section 4, and conclusions and future work will wrap up the paper.

\section{Related work}
Prior to the development of signatures for anti-malware products, the two main tasks that have to be carried out within the scope of malware analysis are malware detection, and malware classification. While the goal of malware detection mechanisms is to catch the malware in the wild, malware classification systems assign each sample to the correct malware family. These systems can be roughly divided into two groups, based, respectively, on dynamic or static analysis.
\\
\textbf{Dynamic analysis.} Researchers have put a lot of effort in proposing behaviour-based malware detection methods that capture the behavior of the program at runtime. One way to observe the behavior of a program is to monitor the interactions of the program with the operating system through the analysis of the API calls \cite{Willems:2007:CWSandbox, Rieck:2008:LCM}. In order to devise an effective and more robust system, some approaches considered additional semantic information like the sequence of the API calls \cite{Ahmadi:2013:MDSP}, and the use of graph representations \cite{Kolbitsch:2009:EEM,Fredrikson:2010:SNM,karbalaie:2012:semantic}. These approaches monitor the program's behaviour by analyzing the temporal order of the API calls, and the effect of API calls on registers \cite{Ghiasi:DVSA:2015}, or by extracting a behavioural graph based on the dependency between API call parameters. Additionally, in contrast to the above program-centric detection approaches, some proposals address the issue by a global, system-wide approach. For example, Lanzi et al. \cite{Lanzi:2010:AUS} proposed an access activity model that captures the generalized interactions of benign applications with operating system resources, such as files and the registry, and then detects the malware  with very a very low false positive rate. 
A recent survey on 36 research papers on dynamic analysis techniques \cite{Rossow:2012:survey} pointed out that the common shortcomings of dynamic analysis techniques are the problematic and somewhat obscure assumptions regarding the use of execution-driven datasets, and the lack of details and motivation on the security precautions that have been taken during the experimental phase. Moreover, recent malware is shipped with  dynamic anti-analysis defenses that hide the malicious behaviour in the case a dynamic analysis environment is detected \cite{Qiu:2014:FUD} and the lack of code coverage, as dynamic analysis is not designed to explore all or, at least, multiple execution paths of an executable \cite{Moser:2007:EME}.
\\
\textbf{Static analysis.} On the other hand, static approaches perform the analysis without actually executing the program. The research literature exhibits a large variety of static analysis methods. SAFE \cite{Christodorescu:2003:SAE} and SAVE \cite{Sung:2004:SAV} have been among the most influential approaches in heuristic static malware detection, as these works inspired many researchers in this area. The above two works proposed the use of different patterns to detect the presence of malicious content in executable files. Since that time, a large variety of techniques have been explored based on different malware attributes, such as the header of the PE, the body of the PE, or both of them. Analysis is further carried out either directly on the bytecode \cite{Tabish:2009:MDU, Nissim:2014:NovelActive}, or by disassembling the code and extracting opcodes and other relevant detailed information on the content of the program \cite{Santos:2013:opcode}. 
The main issue in static analysis is coping with packing and obfuscation. Recently, some paper addressed this issue by proposing a generic approach for the automatic deobfuscation of obfuscated programs without making any assumption about the obfuscation technique \cite{Yadegari:2015:generic}. Static techniques have been also employed to assess if a malware detected in the wild is similar to a previously-seen variant, without actually performing the costly task of unpacking \cite{Jacob:2012:SPF, Narouei:2015:DLLMiner}.
\\
All of the malware detection and malware classification systems rely on the extraction of either static or dynamic features. So, basically, the same features used for malware detection are used for malware classification purposes. As this paper focuses on malware classification based on the extraction of static features, Table~\ref{tab:RelWorks} summarize the prominent static techniques tailored to both the detection and the classification of PE malware designed for MS Windows systems. As far as the experiments reported in the literature  have been performed on different datasets, we haven't reported the related performances, as a comparison of the attained accuracy would have not been fair. Table~\ref{tab:RelWorks} shows, in the \texttt{type} column, if the paper is related to malware detection or classification. The column \texttt{feature} shows if the features are extracted from the PE header or from the PE body. Finally, the \texttt{structure} column reports on the extraction of any complex features, related, for example, to a relationship or a dependency among PE elements.

\begin{table}
\centering
\caption{Static analysis techniques on Windows malware.}
\label{tab:RelWorks}       
\resizebox{!}{2.2cm}{
\begin{tabular}{clcclll}
\hline\noalign{\smallskip} 
\multirow{2}{*}{Year} & \multirow{2}{*}{Authors} & \multicolumn{2}{c}{Type} & \multicolumn{2}{c}{Features} & \multirow{2}{*}{Structure} \\ 
& & \textbf{Det} & \textbf{Class} &\textbf{Header} & \textbf{Body} \\
\noalign{\smallskip}\hline\noalign{\smallskip}
2008 & Ye et al. \cite{Ye:2008:IntMal} & \checkmark &   & API & $-$ & Itemset \\
2009 & PE-Miner \cite{Shafiq:2009:PE-Miner} & \checkmark &   & STC & STC & $-$ \\
2009 & Tabish et al. \cite{Tabish:2009:MDU} & \checkmark &   & BYT & BYT & N-gram \\
2009 & Griffin et al. \cite{Griffin:2009:AGS} & \checkmark &   & BYT & BYT & Sequence \\
2009 & Hu et al. \cite{Hu:2009:LMI} & \checkmark &  & $-$ & FC & Graph \\
2010 & Sami et al. \cite{Sami:2010:MDB} & \checkmark &   & API & $-$ & Itemset \\
2011 & Nataraj et al. \cite{Nataraj:2011:MalwareImages} &  & \checkmark & BYT & BYT & $-$  \\
2012 & Jacob et al. \cite{Jacob:2012:SPF} &  & \checkmark & STC & BYT & N-gram  \\
2013 & Santos et al. \cite{Santos:2013:opcode} & \checkmark &   & $-$ & OP & Sequence \\
2014 & Nissim et al. \cite{Nissim:2014:NovelActive} & \checkmark &   & BYT & BYT & N-gram \\
2015 & DLLMiner \cite{Narouei:2015:DLLMiner} & \checkmark & \checkmark & DLL & $-$ & Tree \\
\noalign{\smallskip}\hline
\noalign{\smallskip}
\end{tabular}
}
\begin{flushleft}
\texttt{API}: Application Programming Interface \\
\texttt{BYT}: Byte code, \texttt{FC}: Function Call \\
\texttt{STC}: Structural features, \texttt{OP}: Operation code \\
\end{flushleft}
\end{table}

\section{System architecture}
As this paper focuses on malware classification, the most relevant issue is related to the choice of the features that will be used to represent each malware sample for the classification task. Our approach was guided by the rationale that to attain accurate and fast classification results, so we should integrate different types of features, such as content-based features as well as structural features.

\subsection{Malware representation}
\label{sec:malrep}
Before entering into the details of the features that we extracted for the classification task, we will briefly review the different ways in which a malware sample can be represented. Two common representations of a malware sample are by the hex view, and the assembly view.
The hex view represents the machine code as a sequence of hexadecimal digits, which is the accumulation of consecutive 16-bytes words, like in the following representation:
\\\\
\texttt{{\small 004010D0 8D 15 A8 80 63 00 BF 55 70 00 00 52 FF 72 7C 53}}
\\\\
The first value represents the starting address of these machine codes in the memory, and each value (byte) bears a meaningful element for the PE, like instruction codes or data. 
\\
The task of disassembling a binary executable into its sequence of assembly instructions can be performed by two main techniques, namely by the linear sweep algorithm, and the recursive traversal algorithm \cite{Schwarz:2002:DEC}. Although neither approach is absolutely precise, the recursive approach is usually far less susceptible to mistakes than the linear sweep algorithm because the code is disassembled according to the jump and branch instructions. The Interactive Disassembler (IDA) \cite{IDA} tool is one of the most popular recursive traversal disassembler, which performs automatic code analysis on binary files using cross-references between code sections, knowledge of parameters of API calls, and other information. For example, IDA interprets the aforementioned byte sequence as shown in Figure~\ref{fig:assemblerview}.

\begin{figure}[htp]
\centering
\NumTabs{2}
\begin{tcolorbox}
\begin{tiny}
.text:00635CD0 8D 15 A8	80 63 00				\tab{}	       lea     edx, unk\_6380A8 \\
.text:00635CD6 BF 55 70	00 00				\tab{}		       mov     edi, 7055h \\
.text:00635CDB 52				\tab{}			       push    edx \\
.text:00635CDC FF 72 7C				\tab{}			       push    dword ptr [edx+7Ch] \\
.text:00635CDF 53		\tab{}		push    ebx	 \\
\tab{} \vspace{-5mm}
\end{tiny}
\end{tcolorbox}
\caption{Assembly view.}
\label{fig:assemblerview}
\end{figure}

\subsection{Features}
For accurate and fast classification, we propose to extract features both from the hex view, and from the assembly view to exploit complementary information brought by these two representations. These complementary information are usually related to the essence of maliciousness, like obfuscation, and the experimental results will show how the combination of information from the two views can help improving the effectiveness of the whole system. In the following subsections we provide details on each feature that has been used and the reasoning of selecting them.
\\
It is worth to point out the reason why we are not considering features extracted from the PE header. While it is well known that the PE header can be a rich source of information, the task at hand is more challenging as the PE header is not available, according to the rules of the Microsoft challenge that provided the dataset used in this paper.

\subsubsection{Hex dump-based features}

\begin{figure*}
\begin{subfigure}{.45\textwidth}
  \centering
  \includegraphics[width=.3\textwidth]{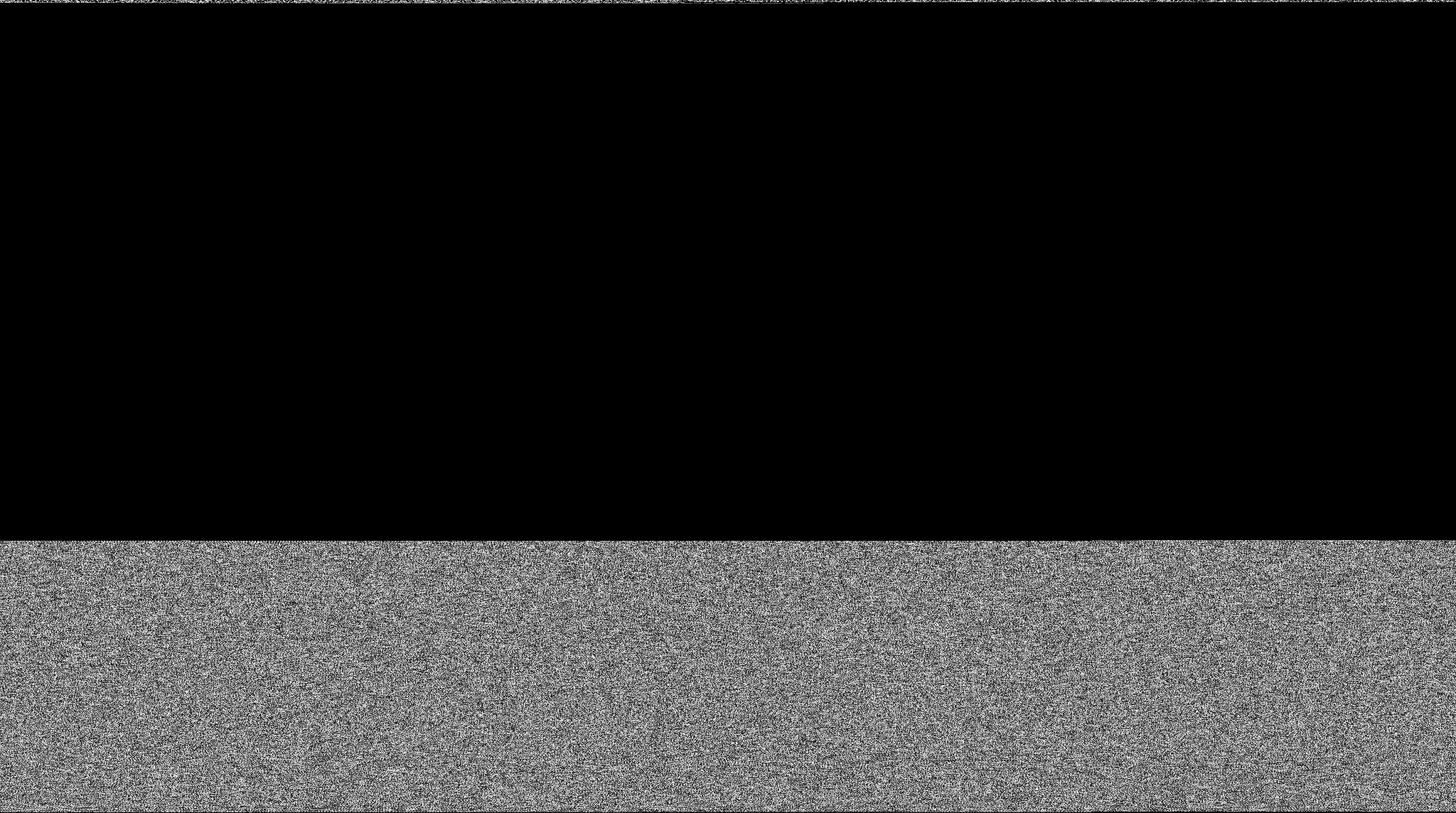}\hfill
\includegraphics[width=.3\textwidth]{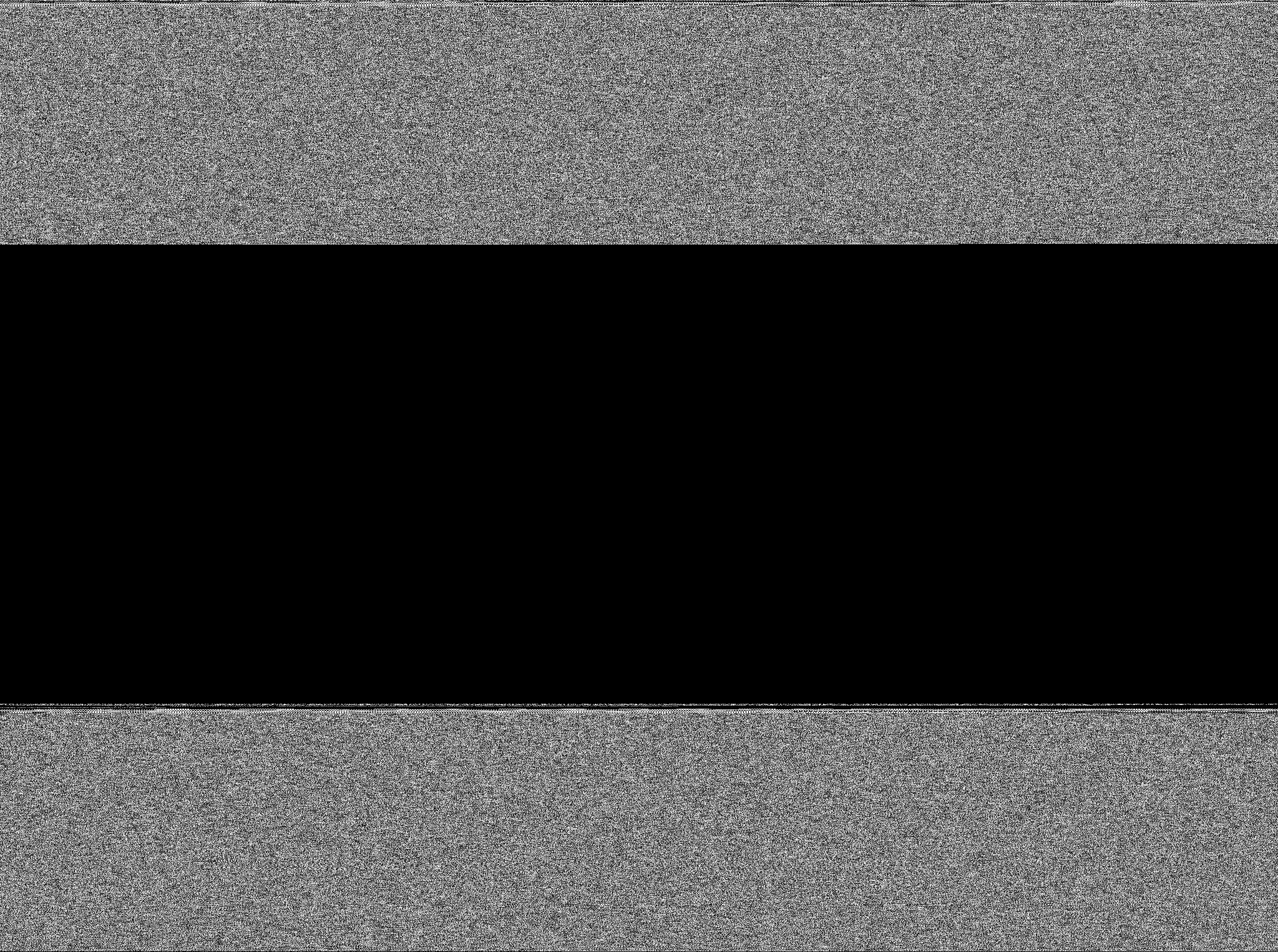}\hfill
\includegraphics[width=.3\textwidth]{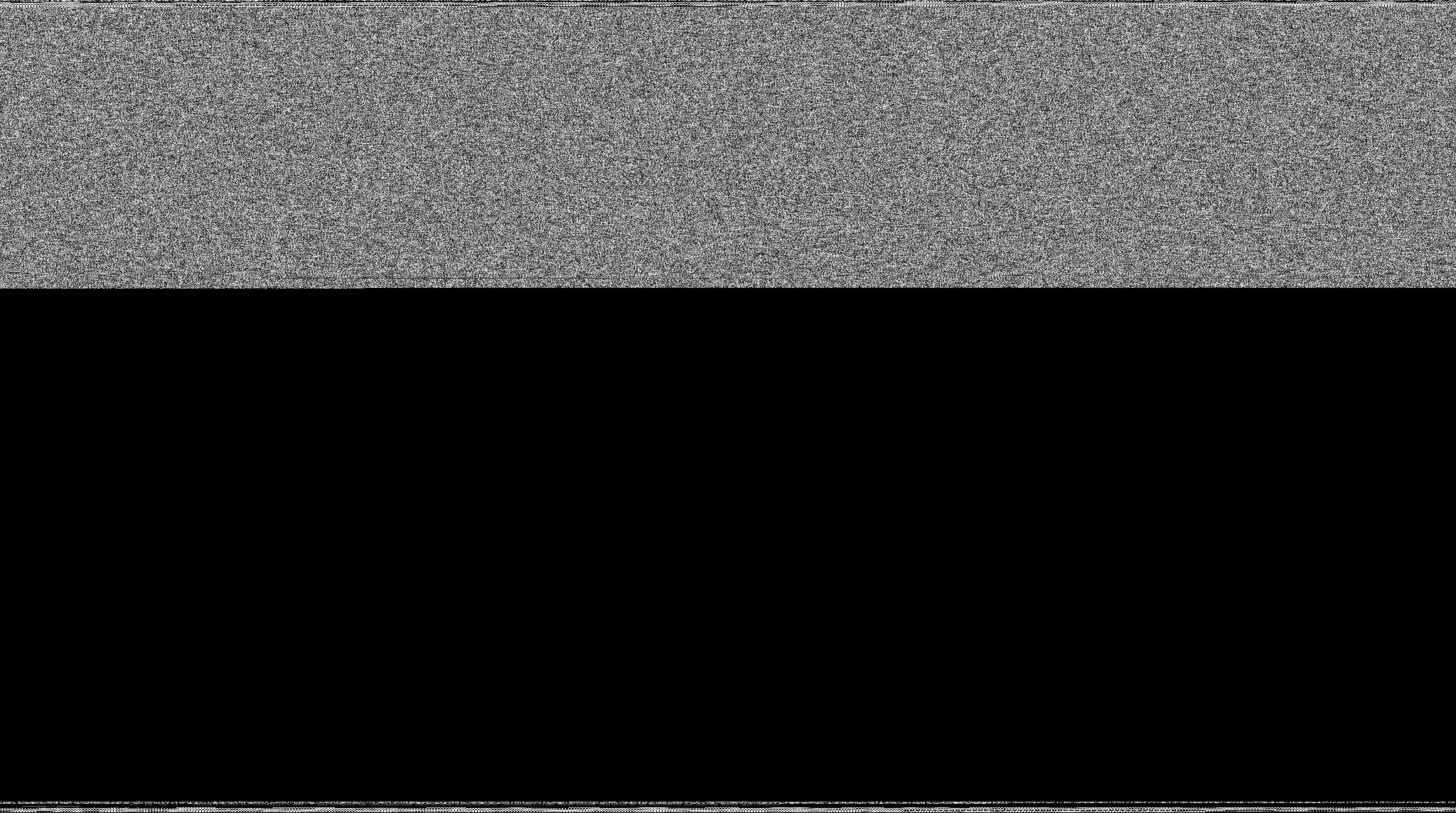}
\caption{Three malware samples in class 3.}
  \label{fig:class3}
\end{subfigure}%
\hspace{5mm}
\begin{subfigure}{.45\textwidth}
  \centering
  \includegraphics[width=.3\textwidth]{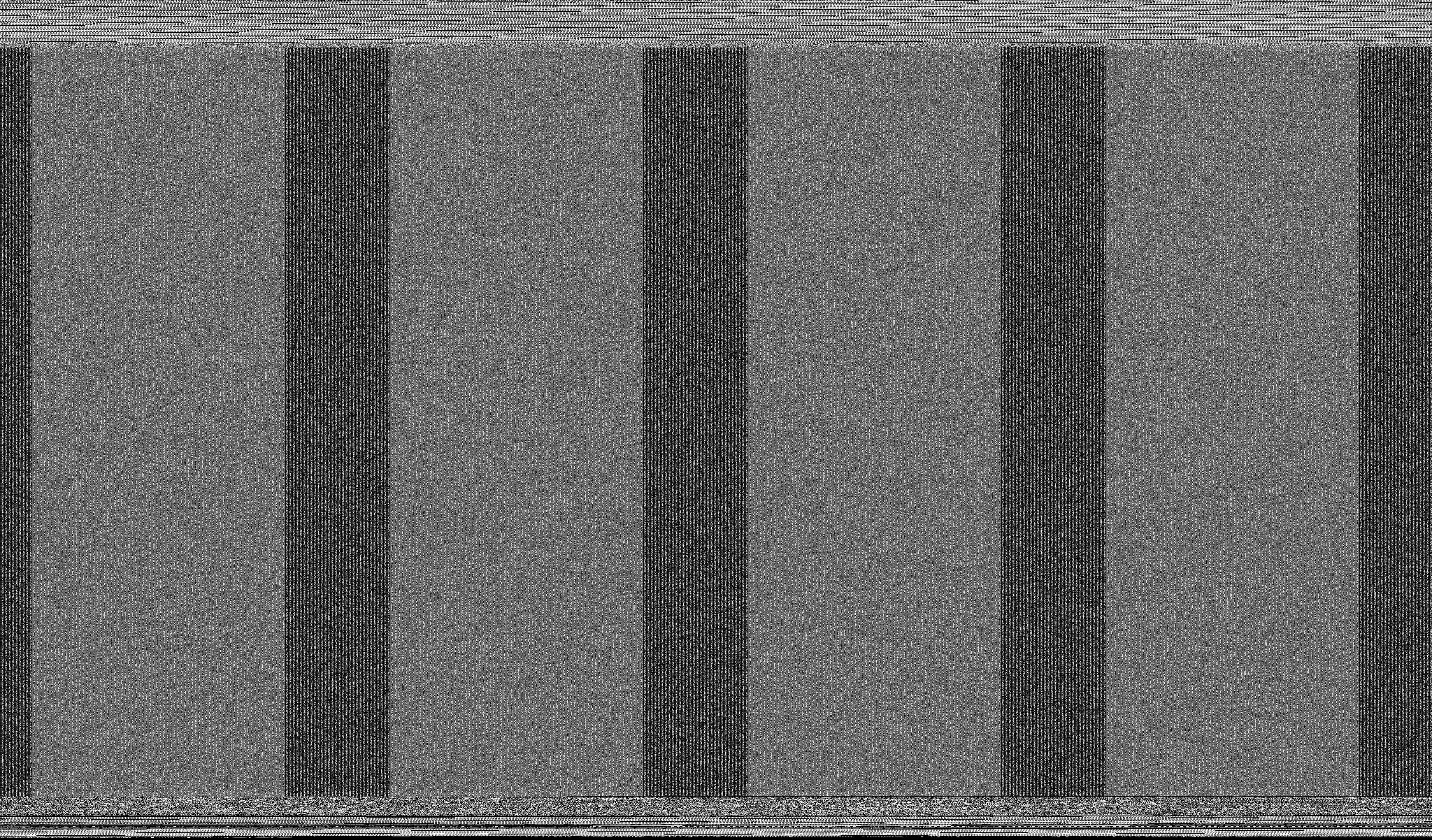}\hfill
\includegraphics[width=.3\textwidth]{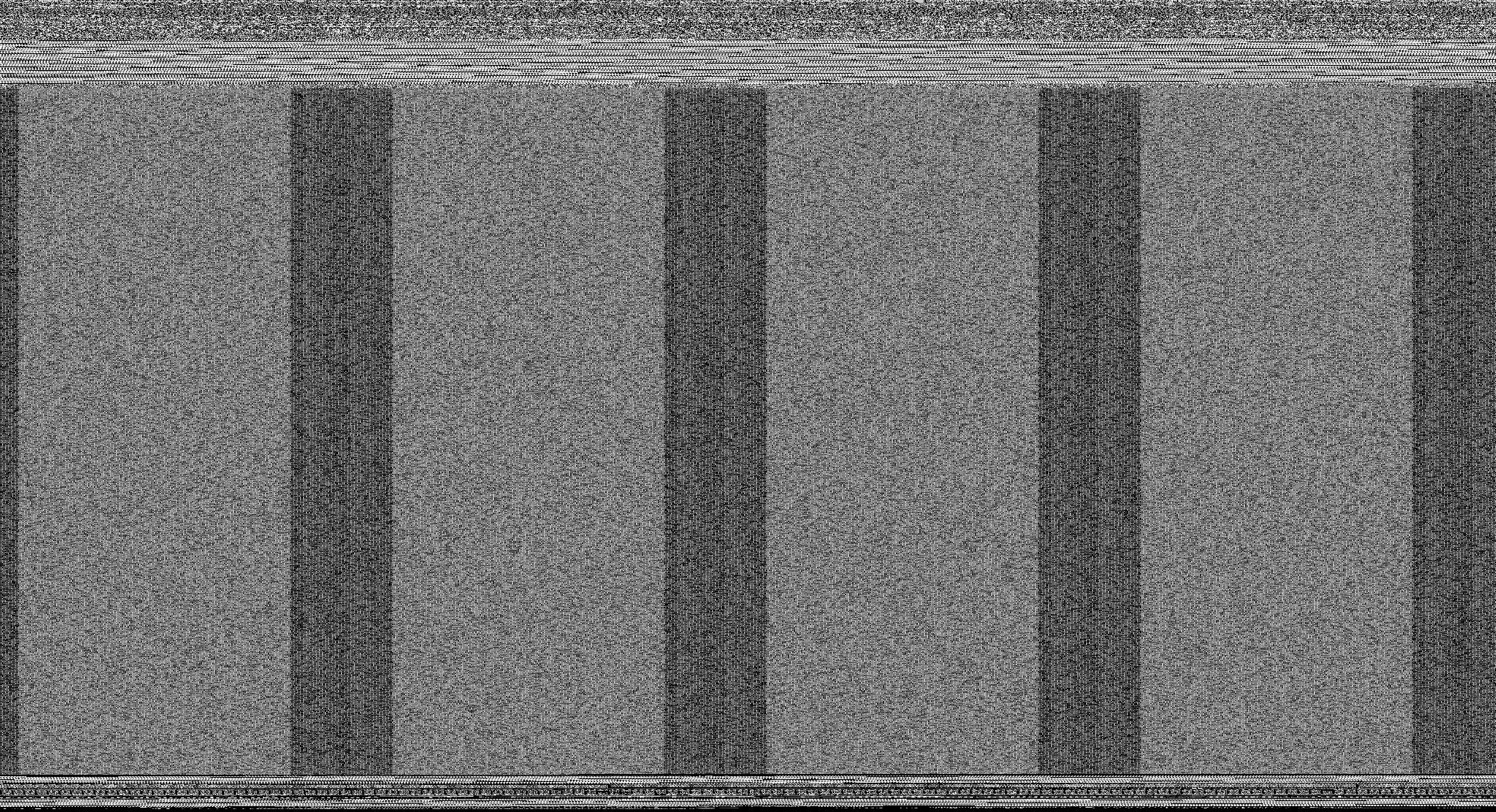}\hfill
\includegraphics[width=.3\textwidth]{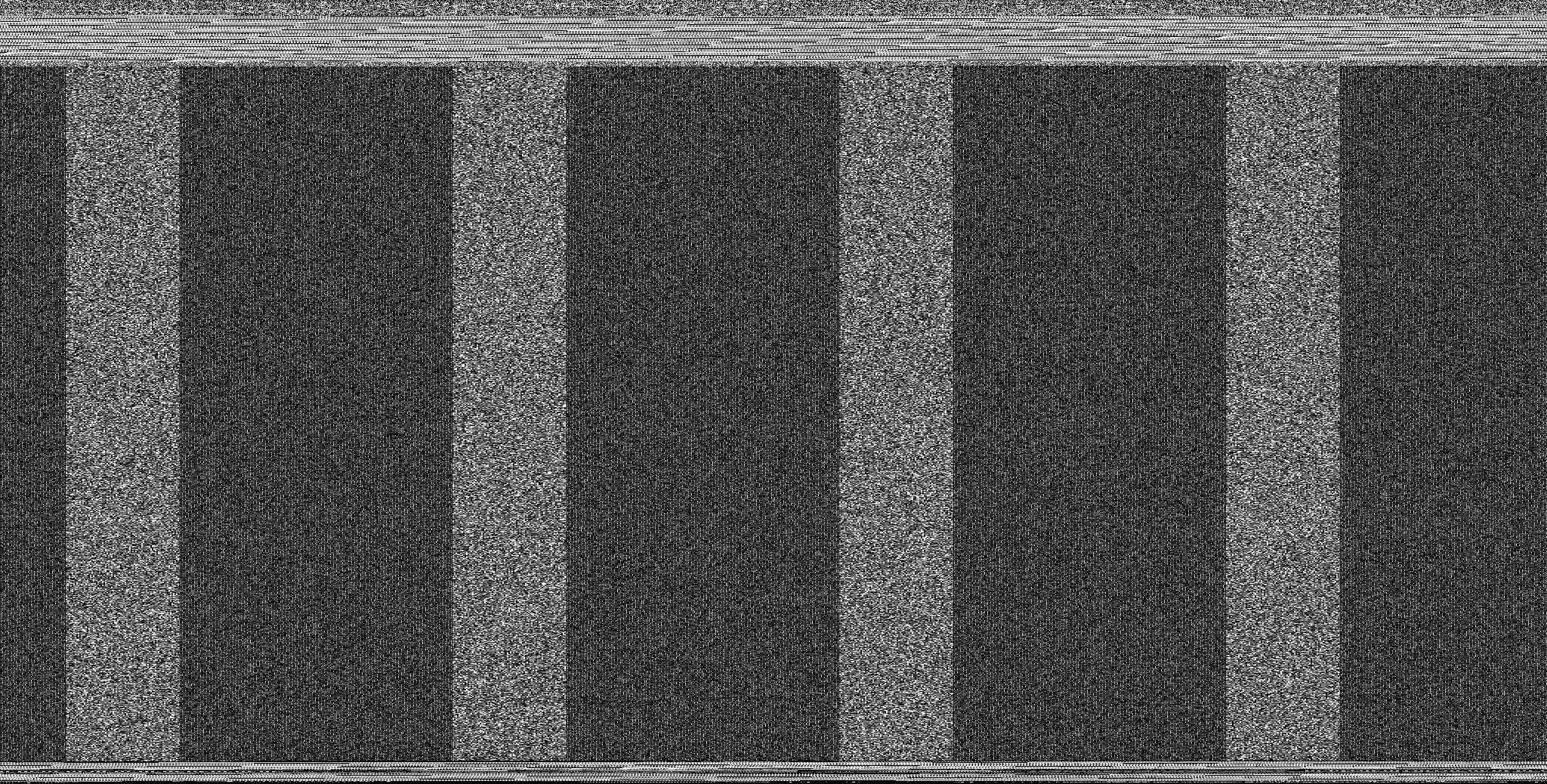}
\caption{Three malware samples in class 2.}
  \label{fig:class2}
\end{subfigure}
\\
\begin{subfigure}{.45\textwidth}
  \centering
\includegraphics[width=.3\textwidth]{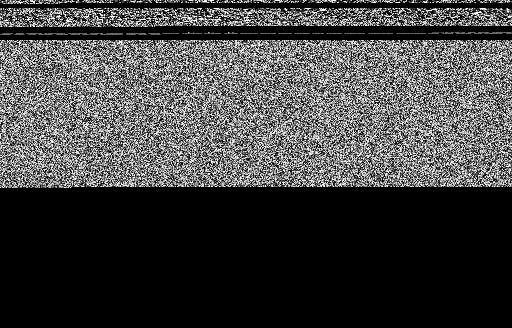}\hfill
\includegraphics[width=.3\textwidth]{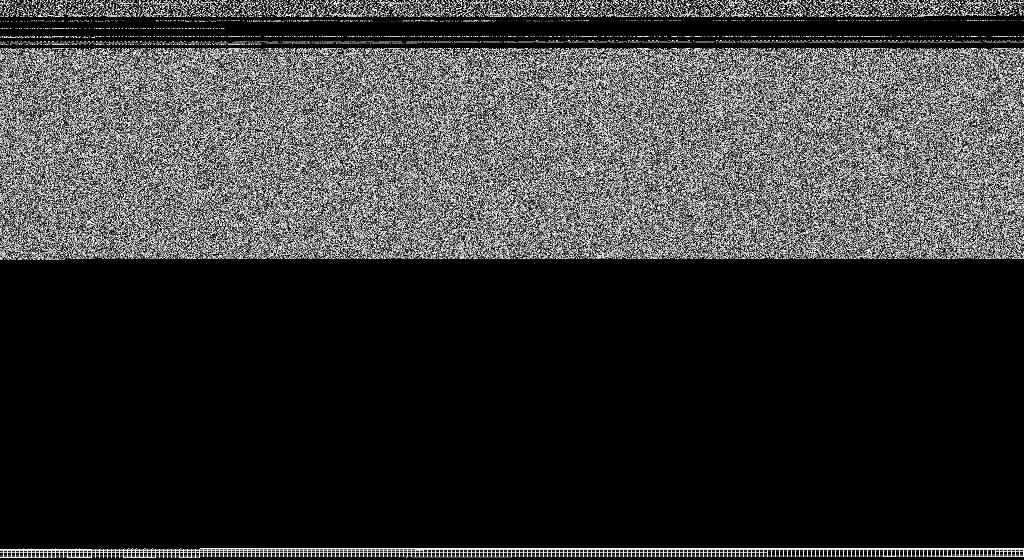}\hfill
\includegraphics[width=.3\textwidth]{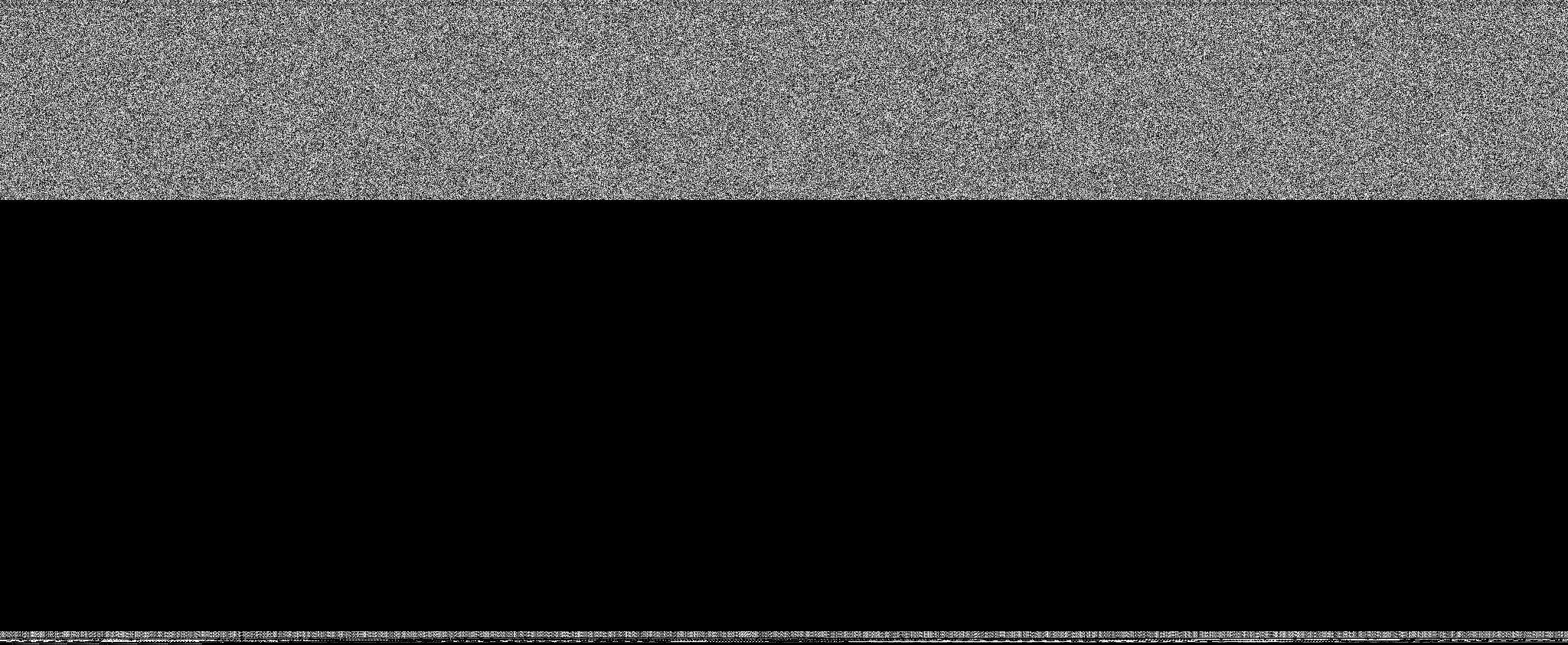}
\caption{Three almost similar images in different classes of 4, 5, 6.}
  \label{fig:class456}
\end{subfigure}%
\hspace{5mm}
\begin{subfigure}{.45\textwidth}
  \centering
\includegraphics[width=.3\textwidth]{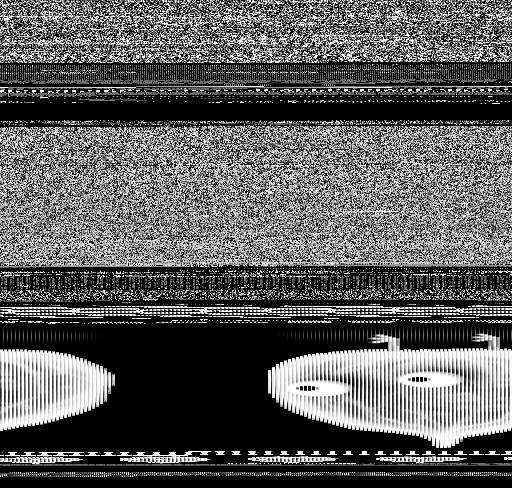}\hfill
\includegraphics[width=.3\textwidth]{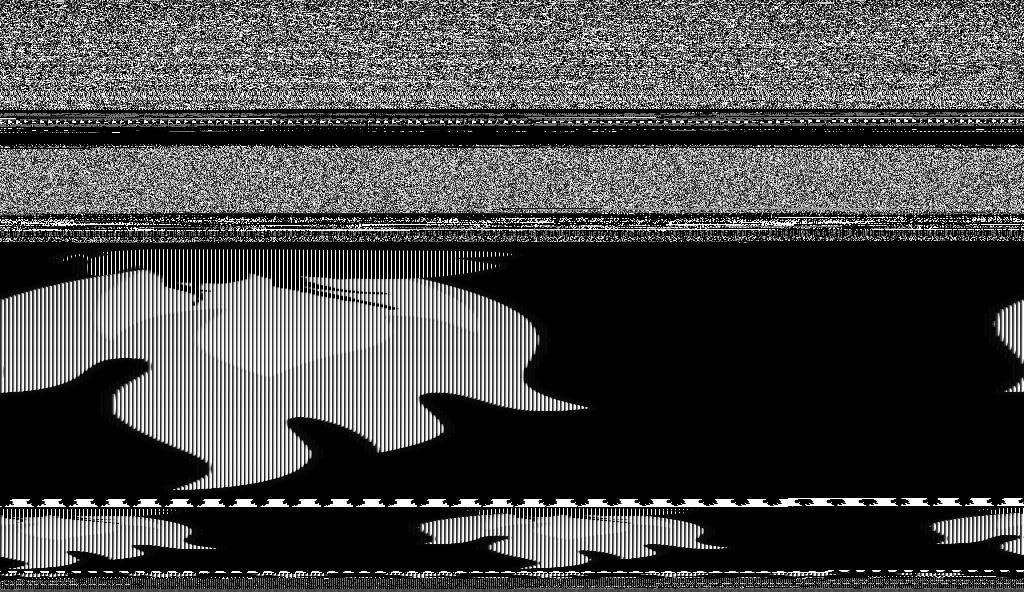}\hfill
\includegraphics[width=.3\textwidth]{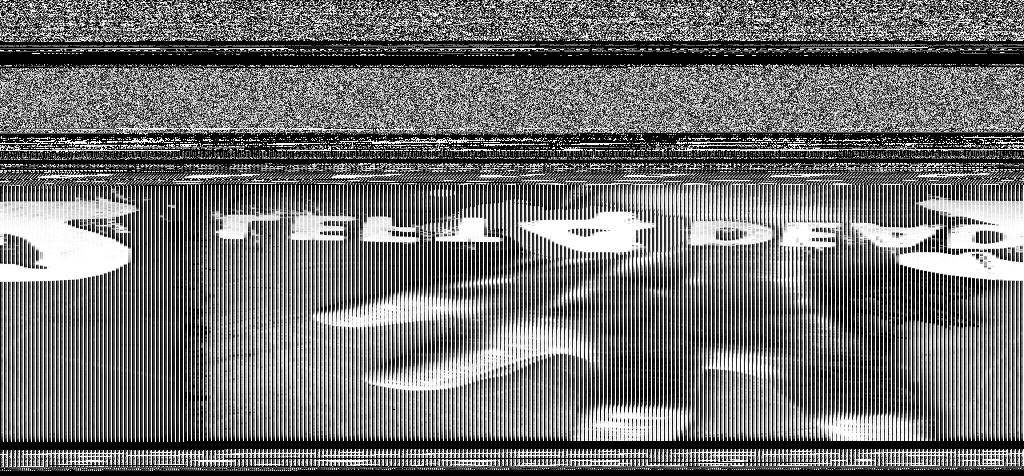}
\caption{Some images embedded in malware.}
  \label{fig:rsrc}
\end{subfigure}
\caption{Image representation of malware samples.}
\label{fig:Image}
\end{figure*}

\begin{enumerate}
\item \textbf{\texttt{N-gram:}}
\\
A N-gram is a contiguous sequence of n items from a given sequence. N-gram is intensively used for characterizing sequences in different areas, e.g. computational linguistics, and DNA sequencing. The representation of a malware sample as a sequence of hex values can be effectively described through n-gram analysis to capture beneficial information about the type of malware. Each element in a byte sequence can take one out of 257 different values, i.e., the 256 byte range, plus the special $??$ symbol. The ``??" symbol indicates that the corresponding byte has no mapping in the executable file, namely the contents of those addresses are uninitialized within the file. This value can be discarded as, from an experimental point of view, it turned out that better results are achieved by taking into account just the 256 symbols.
Examples of N-gram analysis include 1-gram (\textbf{1G}) features, which represent just the byte frequency, and thus are described with a 256-dimensional vector, and 2-gram features, which measure the frequency of all 2-byte combinations, thus having dimension of $256^2$. As far as low computational complexity is concerned in our assumption, 1-gram is just considered in the experiments.
\item \textbf{\texttt{Metadata:}} 
\\
We extract the following metadata features (\textbf{MD1}), namely, the size of the file, and the address of the first bytes sequence. The address is an hexadecimal number, and we converted it to the corresponding decimal value for homogeneity with the other features values.
\item \textbf{\texttt{Entropy:}} 
\\
Entropy (\textbf{ENT}) is a measure of the amount of \textit{disorder}, and can be used to detect the possible presence of obfuscation \cite{Lyda:2007:Entropy, Baysa:2013:Ent}. Entropy is computed on the byte-level representation of each malware sample and the goal is to measure the \textit{disorder} of the distribution of bytes in the bytecode as a value between 0 (Order) and 8 (Randomness). First, we apply the sliding window method to represent the malware as a series of entropy measures E = ${e_i: i = 1, ..., N}$, where $e_i$ is the entropy measured in each window, and N is the number of windows, and then the entropy is calculated using the Shannon's formula:

\begin{equation} 
\label{eq_entropy}
e_{i}= -\sum_{j=1}^{m}p\left ( j \right )\log_{2}p\left ( j \right )
\end{equation}

where p(j) is the frequency of byte j within window i, and m is a number of distinct bytes in the window. Then, we consider statistics of entropy sequences obtained using the sliding window method, that is, we calculate the entropy for each window of 10000 bytes and then we consider a number of statistical measures like quantiles, percentiles, mean, and variance of the obtained distribution. In addition, we compute the entropy of all the bytes in a malware.

\item \textbf{\texttt{Image representation:}}
\\
An original way to represent a malware sample is to visualize the byte code by interpreting each byte as the gray-level of one pixel in an image \cite{Nataraj:2011:MalwareImages}. As shown in Figure~\ref{fig:Image}, the resulting images have very fine texture patterns (e.g. see Figure~\ref{fig:class3}, and Figure~\ref{fig:class2}), that can be used as visual signatures for each malware family. Although matching visual patterns need a huge processing time, some features that describe the textures in an image \cite{Mahotas} such as the Haralick features (\textbf{IMG1}), or  the Local Binary Patterns features (\textbf{IMG2}) can be efficient and quite effective for the malware classification task. The representation of malware as images may sometimes cause problems, as in the case shown in Figure~\ref{fig:class456}, where the texture patterns of the two images are almost similar, even if the two malware samples that are represented belong to different classes. In addition, we have to take into account the case in which the resources (\texttt{.rsrc}) section of a PE file contains image files (e.g. see Figure~\ref{fig:rsrc}). As the same image files can be used as resources for different malware families, the extracted image patterns from these part of the malware may produce false positives. As far as the .rsrc section may not be always in the same position within a PE file, removing those parts from our analysis was not an easy task. Therefore, as these features are used in conjunction with other feature, we consider the texture patterns computed over the whole image.

\item \textbf{\texttt{String length:}}
\\
We extract possible ASCII strings from each PE using its hex dump. Since this method extracts a lot of garbage along with actual strings, the usage of string features directly is inappropriate. Consequently, to reduce noise and to avoid overfitting, only histograms related to the distribution of length of strings (\textbf{STR}) is used.

\end{enumerate}

\subsubsection{Features extracted from disassembled files}

\begin{enumerate}
\item \textbf{\texttt{Metadata:}} 
\\
After disassembling, we computed the size of the file, and the number of lines in the file, and included these features within the Metadata category (\textbf{MD2}).

\item \textbf{\texttt{Symbol:}}
\\
The frequencies of the following set of symbols (\textbf{SYM}), \texttt{-}, \texttt{+}, \texttt{*}, \texttt{]}, \texttt{[}, \texttt{?}, \texttt{@},  
are taken into account as a high frequency of these characters is typical of code that has been designed to evade detection, for example by resorting to indirect calls, or dynamic library loading. In indirect calls, the address of the callee is taken from the memory/register. Although the implementation of calls depends both on the architecture, and on the optimal decision of compiler, indirect calls may reveal some information on data location obfuscation \cite{Moser:2007:limits}. Dynamic library loading is another mechanism where an executable file loads a library into memory at runtime, and accesses its functions based on their address, so that static analyzers cannot capture the name of the imported functions.

\item \textbf{\texttt{Operation Code:}}
\\
Operation codes (\textbf{OPC}) are the mnemonic representation of machine code, which symbolize assembly instruction. The full list of x86 instruction set is large and complex, so we select a subset of 93 operation codes based either on their commonness, or on their frequent use in malicious applications \cite{Bilar:2006:bh}, and measure the frequency of them in each malware sample. While instruction replacement techniques can be used to evade detection \cite{Christodorescu:2005:Semantics}, their effects on malware classification tasks is limited, both for its rare use, and, consequently, for its negligible contribution to the computation of the statistics.

\item \textbf{\texttt{Register:}}
\\
Most of the processor registers in x86 architecture are used for dedicated tasks, but in some cases register renaming is used to make the analysis harder \cite{Christodorescu:2005:Semantics}. Consequently, the frequency of use of the registers (\textbf{REG}) can be a useful helper for assign a malware sample to one family, as the experiments will show.

\item \textbf{\texttt{Application Programming Interface:}}
\\
We also measure the frequency of use of Windows Application Programming Interfaces (\textbf{API}). As far as the total number of APIs is extremely large, considering them all would bring little or no meaningful information for malware classification. Consequently, we restricted our analysis to the top 794 frequent APIs used by malicious binaries based on an analysis on near 500K malware samples \cite{bnxnet}. This feature category is discriminative for a subset of malware samples, because some samples might contain any API call because of packing, while some other samples might load some of its APIs by resorting to dynamic loading through the \texttt{LoadLibrary} API. For example, the sample with hash code \texttt{00yCuplj2VTc9ShXZDvnxz} was packed with aspack\footnote{http://www.aspack.com/}, and it does not contain any API call, and most of the disassembled code just contains data define instructions like \texttt{db} (see Figure~\ref{fig:db}) and \texttt{dd} (see Figure~\ref{fig:dd}).

\begin{figure}[htp]
\centering
\NumTabs{2}
\begin{tcolorbox}
\begin{tiny}
DATA:0042F259 E1				\tab{}			      db 0E1h ;	\'{a}	\\
DATA:0042F25A 36				\tab{}			      db  36h ;	6	\\
DATA:0042F25B 4E				\tab{}			      db  4Eh ;	N	\\
DATA:0042F25C 12				\tab{}			      db  12h			\\
DATA:0042F25D 45				\tab{}			      db  45h ;	E	\\
DATA:0042F25E 0B				\tab{}			      db  0Bh			\\
DATA:0042F25F 4A				\tab{}			      db  4Ah ;	J	\\
DATA:0042F260 43				\tab{}			      db  43h ;	C	\\
DATA:0042F261 6A				\tab{}			      db  6Ah ;	j	\\
DATA:0042F262 18				\tab{}			      db  18h			\\
DATA:0042F263 DB				\tab{}			      db 0DBh ;	\^{U}	\\
DATA:0042F264 A7				\tab{}			      db 0A7h ;	\S{}		\\
\tab{} \vspace{-5mm}
\end{tiny}
\end{tcolorbox}
\caption{A part of \texttt{00yCuplj2VTc9ShXZDvnxz} (Packed, Changing section name); The sample contains no API call, and just few assembly instructions.}
\label{fig:db}
\end{figure}

\begin{figure*}[htp]
\begin{tcolorbox}
\begin{tiny}
.aspack:004BFA2C 20 20 20 00 34	34 34 00 56 56 56 00 0B	0B 0B 7B+ \tab{dd 202020h, 343434h, 565656h, 7B0B0B0Bh, 0FF292929h, 0FC282828h}\\
.aspack:004BFA2C 29 29 29 FF 28	28 28 FC 2D 2D 2D FE 2C	2B 2A FF+	 \tab{dd 0FE2D2D2Dh,	0FF2A2B2Ch, 0FF060504h,	0FF824B03h, 0FFE89325h}\\
.aspack:004BFA2C 04 05 06 FF 03	4B 82 FF 25 93 E8 FF 40	A6 F5 FF+	 \tab{dd 0FFF5A640h,	0FFFAA737h, 0FFFCAC37h,	0FFFBAD2Eh, 0FFFBAC23h}\\
.aspack:004BFA2C 37 A7 FA FF 37	AC FC FF 2E AD FB FF 23	AC FB FF+	 \tab{dd 0FFFBAE1Dh,	0FFFAAD16h, 0FFFBB014h,	0FFFAB71Eh, 0FFFFBE14h}\\
.aspack:004BFA2C 1D AE FB FF 16	AD FA FF 14 B0 FB FF 1E	B7 FA FF+		 \tab{dd 0FF8B6A34h,	0FF282E3Ah, 0FF363634h,	0FF323134h, 0FFDE9C14h}\\
.aspack:004BFA2C 14 BE FF FF 34	6A 8B FF 3A 2E 28 FF 34	36 36 FF+		 \tab{dd 0FFFEBC16h,	0FFF7AA10h, 0FFF9AB13h,	0FFFCB221h, 0FFFBAC21h}\\
.aspack:004BFA2C 34 31 32 FF 14	9C DE FF 16 BC FE FF 10	AA F7 FF+		 \tab{dd 0FFFAAB2Bh,	0FFFCAE36h, 0FFFBA835h,	0FFF9A941h, 0FFF19E32h}\\
.aspack:004BFA2C 13 AB F9 FF 21	B2 FC FF 21 AC FB FF 2B	AB FA FF+		 \tab{dd 0FFC2720Ah,	0FF321D02h, 0FE000004h,	0FE0A0909h, 0FC080808h} \\
.aspack:004BFA2C 36 AE FC FF 35	A8 FB FF 41 A9 F9 FF 32	9E F1 FF+	 \tab{dd 0FE060606h,	0E0101010h, 181B1B1Bh, 10101h, 202020h} \tab{} \vspace{-5mm}
\end{tiny}
\end{tcolorbox}
\caption{A part of \texttt{00yCuplj2VTc9ShXZDvnxz} (Packed, Changing section name); The sample contains no API call, and just few assembly instructions.}
\label{fig:dd}
\end{figure*}

\item \textbf{\texttt{Section:}}
\\
A PE consists of some predefined sections like \texttt{.text}, \texttt{.data}, \texttt{.bss}, \texttt{.rdata}, \texttt{.edata}, \texttt{.idata}, \texttt{.rsrc}, \texttt{.tls}, and \texttt{.reloc}. Because of evasion techniques like packing, the default sections can be modified, reordered, and new sections can be created. We extract different characteristics from sections (\textbf{SEC}), which are listed in Table~\ref{tab:SEC}. In Section~\ref{sec:fi} we will point out that this category is the one with the higher influence in the classification performances.

\begin{table}
\centering
\caption{List of features in the SEC category.}
\label{tab:SEC}       
\resizebox{!}{3.0cm}{
\begin{tabular}{cl}
\hline\noalign{\smallskip}
\textbf{Name} & \textbf{Description} \\
\noalign{\smallskip}\hline\noalign{\smallskip}
 section\_names\_.bss & The total number of lines in .bss section  \\
 section\_names\_.data	&	The total number of lines in .data section	\\
 section\_names\_.edata	&	The total number of lines in .edata section	\\
 section\_names\_.idata		&	The total number of lines in .idata section	\\
 section\_names\_.rdata	&	The total number of lines in .rdata section	\\
 section\_names\_.rsrc	&	The total number of lines in .rsrc section	\\
 section\_names\_.text	&	The total number of lines in .text section	\\
 section\_names\_.tls	&	The total number of lines in .tls section	\\
 section\_names\_.reloc &	The total number of lines in .reloc section	\\
 Num\_Sections	&	The total number of sections	\\
 Unknown\_Sections	&	The total number of unknown sections	\\
 Unknown\_Sections\_lines	&	The total number of lines in unknown sections	\\
 known\_Sections\_por	&	The proportion of known sections to the all section\\
 Unknown\_Sections\_por	&	The proportion of unknown sections to the all sections	\\
 Unknown\_Sections\_lines\_por	&	The proportion of the amount of unknown sections to the whole file	\\
 .text\_por	&	The proportion of .text section to the whole file	\\
 .data\_por	&	The proportion of .data section to the whole file	\\
 .bss\_por	&	The proportion of .bss section to the whole file		\\
 .rdata\_por	&	The proportion of .rdata section to the whole file		\\
 .edata\_por	&	The proportion of .edata section to the whole file		\\
 .idata\_por	&	The proportion of .idata section to the whole file		\\
 .rsrc\_por	&	The proportion of .rsrc section to the whole file		\\
 .tls\_por	&	The proportion of .tls section to the whole file		\\
 .reloc\_por	&	The proportion of .reloc section to the whole file	 \\
\noalign{\smallskip}\hline
\end{tabular}
}
\end{table}

\item \textbf{\texttt{Data Define:}}
\\
As shown in Figure~\ref{fig:db} and Figure~\ref{fig:dd}, some malware samples do not contain any API call, and just contain few operation codes, because of packing, In particular, they mostly contain \texttt{db}, \texttt{dw}, and \texttt{dd} instructions, which are used for setting byte, word, and double word respectively. Consequently, we propose to include this novel set of features (\textbf{DP}) for malware classification as it has a high discriminative power for a number of malware families. The full list of features in this category is presented in Table~\ref{tab:DP}.

\begin{table}
\centering
\caption{List of features in the DP category.}
\label{tab:DP}       
\resizebox{!}{2.85cm}{
\begin{tabular}{cl}
\hline\noalign{\smallskip}
\textbf{Name} & \textbf{Description} \\
\noalign{\smallskip}\hline\noalign{\smallskip}
 db\_por & The proportion of \texttt{db} instructions in the whole file  \\
 dd\_por & The proportion of \texttt{dd} instruction in the whole file  \\
 dw\_por & The proportion of \texttt{dw} instruction in the whole file  \\
 dc\_por & The proportion of all \texttt{db}, \texttt{dd}, and \texttt{dw} instructions in the whole file  \\
 db0\_por &  The proportion of \texttt{db} instruction with 0 parameter in the whole file \\ 
 dbN0\_por & The proportion of \texttt{db} instruction with not 0 parameter in the whole file  \\
 dd\_text & The proportion of \texttt{dd} instruction in the text section   \\
 db\_text & The proportion of \texttt{db} instruction in the text section \\
 dd\_rdata  & The proportion of \texttt{dd} instruction in the rdata section  \\
 db3\_rdata & The proportion of \texttt{db} instruction with one non 0 parameter in the rdata section  \\
 db3\_data & The proportion of \texttt{db} instruction with one non 0 parameter in the data section   \\
 db3\_all  &  The proportion of \texttt{db} instruction with one non 0 parameter in the whole file  \\
 dd4  &  The proportion of \texttt{dd} instruction with four parameters	\\
 dd5	& The proportion of \texttt{dd} instruction with five parameters	\\
 dd6	& The proportion of \texttt{dd} instruction with six parameters	\\
 dd4\_all	& The proportion of \texttt{dd} instruction with four parameters in the whole file 	\\
 dd5\_all	& The proportion of \texttt{dd} instruction with five parameters in the whole file	\\
 dd6\_all	& The proportion of \texttt{dd} instruction with six parameters in the whole file	\\
 db3\_idata	&  The proportion of \texttt{db} instruction with one non 0 parameter in the idata section	\\
 db3\_NdNt	&  The proportion of \texttt{db} instruction with one non 0 parameter in unknown sections	\\
 dd4\_NdNt	&  The proportion of \texttt{dd} instruction with four parameters in unknown sections	\\
 dd5\_NdNt	&  The proportion of \texttt{dd} instruction with five parameters in unknown sections	\\
 dd6\_NdNt	&  The proportion of \texttt{dd} instruction with six parameters in unknown sections	\\
 db3\_zero\_all	&  The proportion of \texttt{db} instruction with 0 parameter to \texttt{db} instruction with non 0 parameter	\\
\noalign{\smallskip}\hline
\end{tabular}
}
\end{table}

\item \textbf{\texttt{Miscellaneous:}}
We extract the frequency of 95 manually chosen keywords (\textbf{MISC}) from the disassembled code. Some of these keywords are related to the interpretation of IDA from the code, like 75 adjacent dash-lines which show the border of blocks of PE, and counting them represent the number of blocks in PE. Others are some strings like \texttt{hkey\_local\_machine} which represent the access to a specific path of the Windows registry, and the rest are related to the code like \texttt{dll} which shows the number of imported DLLs. Because of the limitation of the pages of the paper, the full list will be available in our online repository.
\end{enumerate}

\subsection{Feature fusion}
The simplest way for combining feature categories is to stack all the feature categories in a single, long feature vector, and then run a classifier on them. However, it is often in the feature selection process that some of the features turn out to be irrelevant for class discrimination. Including such irrelevant features leads not only to unnecessary computational complexity, but also to the potential decrease of the accuracy of the resulting model. Within the vast literature on feature selection, we focused on two approaches. One approach is the \textit{best subset} selection technique  \cite{James:2014:ISL} that can be summarized as follows. Starting with subsets containing just one feature, a classifier is trained, and the subsets with the highest value of the objective function used to assess the performance (e.g., accuracy, loss functions, etc.) is retained. Then, the process is repeated for any subset containing $f$ features, where $f$ is increased by one at each step so that, for example, all the possible subsets of two features 
$\binom f2=\frac{f(f-1)}{2}$ 
are considered. The other technique that we considered is the \textit{forward stepwise selection} technique which starts with a model containing no feature, and then gradually augments the feature set by adding more features to the model, one by one. This technique for feature selection is computationally more efficient than the \textit{best subset selection} technique because the former just considers $\sum\limits_{i=1}^f (f-k) = \frac{f(f+1)}{2}$ subsets, while the latter considers all $2^f$ possible models, using a greedy approach. 
\\
Based on the above considerations, we implemented an original version of the forward stepwise selection algorithm, where instead of considering one feature at a time, we considered all the subset of features belonging to a \textit{feature category} at a time. At each step, the feature set that produces the minimum value of logloss (see section~\ref{EM} ) will be added to the model. The process stops when adding more features does not decrease the value of logloss.

\subsection{Classification}
As for the feature selection task, over the years a large number of classification techniques have been proposed by the scientific community, and the choice of the most appropriate classifier for a given task is often guided by previous experience in different domains, as well as by trial\&error procedures. However, recently some researchers evaluated the performances of about 180 classifiers arising from different families, using various datasets, and they concluded that random forests and SVM are the two classification mechanisms that have the highest likelihood to produce good performances \cite{Fernandez-Delgado:2014:WNH}. On the other hand, most of the winners in the very recent Kaggle competitions used the XGBoost technique \cite{XGBoost}, which is a parallel implementation of the gradient boosting tree classifier, that in most of the cases produced better performances than those produced by random forests. The XGBoost technique is available as a library, implemented as a parallel algorithm that is fast and efficient, and whose parameters are completely tunable. The high performance and effectiveness of XGBoost is the main motivating reason for using this library for the task at hand. 
In addition, we also used bagging \cite{Breiman:1996:BP} to boost our single model, which is simple, classifier independent, and yet an efficient method to improve the classification quality. More details on the classification technique will be provided in the experimental section. 

\subsection{Evaluation measures}
\label{EM}
The performance in classification has been assessed by using two measures, namely, the accuracy, and the logarithmic loss. The accuracy has been measured as the fraction of correct predictions. As classification accuracy alone is usually not enough to assess the robustness of the prediction, we also measured the logarithmic loss (\texttt{logloss}), 
which is a \texttt{soft} measurement of accuracy that incorporates the concept of probabilistic confidence. It is the \textit{Cross entropy} between the distribution of the true labels and the predicted probabilities. As shown in equation~\ref{eq_logloss}, it is the negative log likelihood of the model,

\begin{equation} 
\label{eq_logloss}
logloss = -\frac{1 }{N}\sum_{i=1}^{N}\sum_{j=1}^{M}y_{ij}log(p_{ij})
\end{equation}

where $N$ is the number of observations, $M$ is the number of class labels, $log$ is the natural logarithm, $y_{ij}$ is $1$ if observation $i$ is in class $j$ and $0$ otherwise, and $p_{ij}$ is the predicted probability that observation $i$ is in class $j$.

\section{Experiments and results}

\subsection{Data}
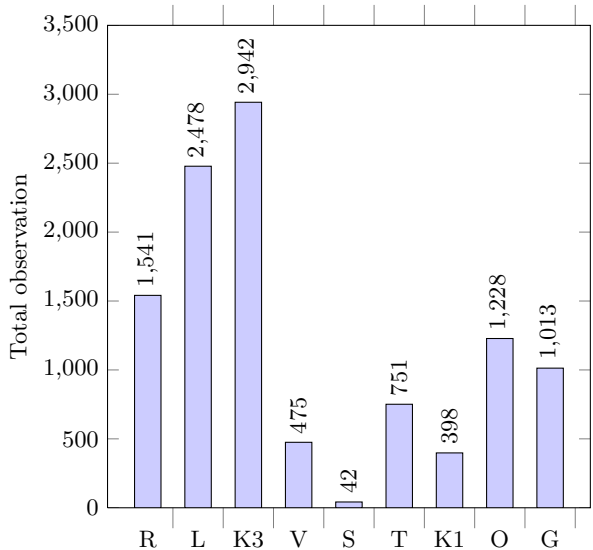
\begin{figure}[h]
\pgfplotstableread{
0 1541
1 2478
2 2942
3 475
4 42
5 751
6 398
7 1228
8 1013
}\dataset
\begin{tikzpicture}
\begin{axis}[ybar,
        width=8cm,
        height=8cm,
        ymin=0,
        ymax=3500,        
        ylabel={Total observation},
        xtick=data,
        xticklabels = {
            \strut R,
            \strut L,
            \strut K3,
            \strut V,
            \strut S,
            \strut T,
            \strut K1,
            \strut O,
            \strut G
        },
        major x tick style = {opacity=0},
        minor x tick num = 1,
        minor tick length=2ex,
        every node near coord/.append style={
                anchor=west,
                rotate=90
        },
        ]
\addplot[draw=black,fill=blue!20, nodes near coords] table[x index=0,y index=1] \dataset; 
\end{axis}
\end{tikzpicture}
\caption{The distribution of data across malware families.}
\label{fig:DataDistr}

\end{figure}

Microsoft released almost half a terabyte of data related to 21741 malware samples, where 10868 samples are used for training, and the rest is for testing. The ID of each malware sample is a 20 characters hash value. The files are from nine different malware families, namely \texttt{Ramnit (R)}, \texttt{Lollipop (L)}, \texttt{Kelihos\_ver3 (K3)}, \texttt{Vundo (V)}, \texttt{Simda (S)}, \texttt{Tracur (T)}, \texttt{Kelihos\_ver1 (K1)}, \texttt{Obfuscator.ACY (O)}, \texttt{Gatak (G)}. The class label of each file is represented by an integer from 1 to 9, where '1' represented the first malware family in the above list, and '9' the last one. There are two files for each malware sample, one containing the hex code, and the other one containing the disassembled code (see Section~\ref{sec:malrep}). Microsoft removed the PE header to ensure file sterility. The distribution of data across the 9 families is shown in Figure~\ref{fig:DataDistr}.

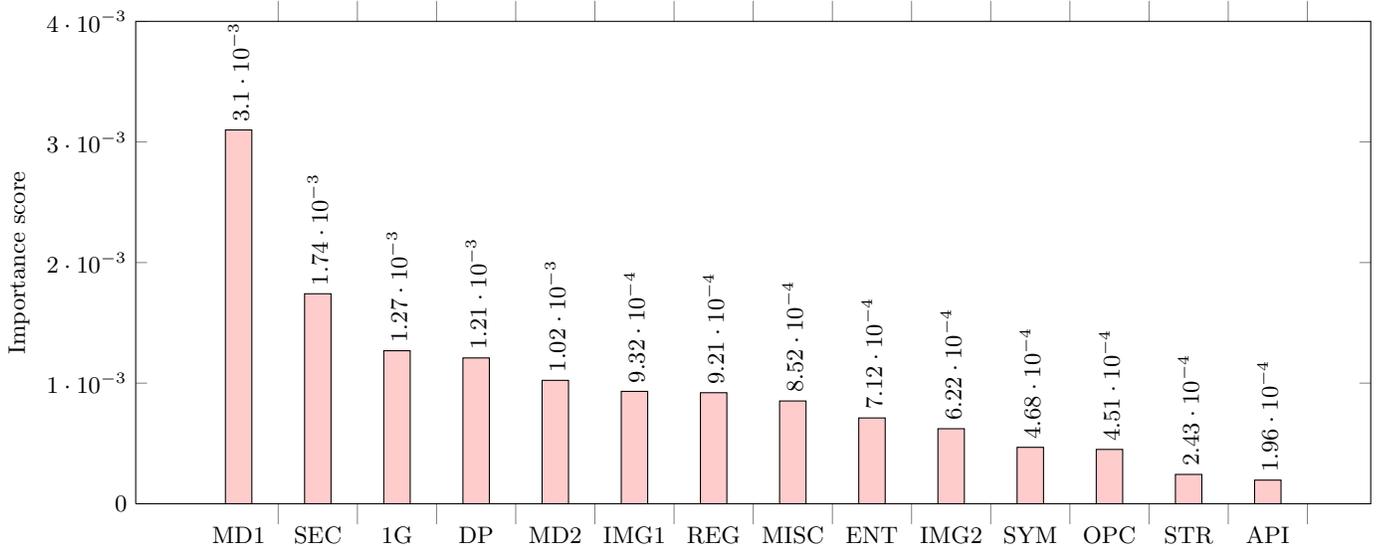
\begin{figure*}
\pgfplotstableread{
0 0.0030995570025
1 0.0017411632601
2 0.00126884359981
3 0.00120900418535
4 0.00102332259639
5 0.000931855124332
6 0.000920723471466
7 0.00085221603012
8 0.000711540876851
9 0.000622157024143
10 0.000467887164445
11 0.000450507373051
12 0.000243195044696
13 0.00019589752256
}\dataset
\begin{tikzpicture}
\begin{axis}[ybar,
        width=18cm,
        height=8cm,
        ymin=0,
        ymax=0.004,        
        ylabel={Importance score},
        xtick=data,
        xticklabels = {
            \strut MD1,
            \strut SEC,
            \strut 1G,
            \strut DP,
            \strut MD2,
            \strut IMG1,
            \strut REG,
            \strut MISC,
            \strut ENT,
            \strut IMG2,
            \strut SYM,
            \strut OPC,
            \strut STR,
            \strut API
        },
        major x tick style = {opacity=0},
        minor x tick num = 1,
        minor tick length=2ex,
        every node near coord/.append style={
                anchor=west,
                rotate=90
        },
        ]
\addplot[draw=black,fill=red!20, nodes near coords] table[x index=0,y index=1] \dataset; 
\end{axis}
\end{tikzpicture}
\caption{Importance of each feature category based on the \textit{mean decrease impurity}.}\label{FI}

\end{figure*}

\subsection{Feature importance}
\label{sec:fi}
Although there is no strict consensus about the meaning of importance, we refer to two common ways to measure the importance of the features when decision tree classifiers are used, i.e., the \textit{mean decrease accuracy}, and the \textit{mean decrease impurity} \cite{cart84}. These two metrics respectively measure the decrease in accuracy or the decrease in impurity\footnote{Gini impurity is a standard decision-tree splitting metric.} associated with each feature. In both cases, the importance of a given feature is proportional to the amount of decrease in accuracy or impurity related to that feature. While in Section~\ref{sub:result} we will discuss the relationship between each feature category and the classification accuracy based on the feature fusion algorithm, in this section we report the importance of the features based on the \textit{mean decrease impurity} to give a better insight on the relevance of each feature category for the attribution of the family to a given malware sample. For this purpose, we used the Random Forest algorithm, and the results are reported in Figure~\ref{FI}. It is worth to point out that Figure~\ref{FI} shows that the two novel structural feature categories that we propose in this paper, namely \texttt{SEC} and \texttt{DP}, are among the top important features that most contribute to the decrease in the impurity of the classification tree.

\subsection{Results}
\label{sub:result}
Table~\ref{results} and Table~\ref{fresults} respectively show the classification performances related to each individual feature category, and the performances related to the combination of feature categories. In particular, Table~\ref{fresults} provides useful information for data analysts to evaluate the trade-off between the number of features used, and the significance of the increase of the classification performances. We proceeded by leveraging on the feature fusion algorithm, by adding one by one the feature category that achieves the lowest logloss on training data. The attained results suggest that the combination of all the feature categories except the \textbf{IMG2} category lead to the lowest logloss on all training data, while the combination of all the feature categories leads to the lowest logloss on training data by employing cross validation. According to these results, we fine tuned the parameters of the XGBoost algorithm on these two feature configurations, as well as for the Bagging technique (see Table~\ref{finalresults}). 
In particular, by adding the external bagging technique, we created a training set with eight times more samples instead of just using the plain training set. We considered all $L$ train samples and sampled $Alpha \times L$ more samples with replacement, where the best value of $Alpha$ was found by grid search and set to one. 

\begin{figure}[htp]
\centering
\includegraphics[width=.40\textwidth]{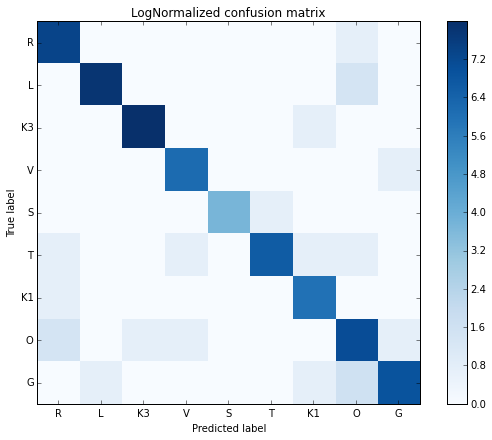}
\caption{The normalized confusion matrix.}
\label{fig:cm2}
\end{figure}

The proposed methodology for the classification of malware allowed achieving a very promising accuracy on the training set of 99.77\%, as well as a very low logloss of 0.0096 on the combination of all categories, and 99.76\% accuracy and 0.0094 logloss on the combination of the best feature categories, based on the outcome of the feature fusion algorithm. The log normalized confusion matrix of the final model is shown in Figure~\ref{fig:cm2}. 

As far as the class labels of the test data were not provided by Microsoft, the only possible way to perform the evaluation on test data is through the submission of the predictions of our model to the competition website. Hence, we ran the experiments on test data and achieved a very low logloss, which is 0.0064 on combination of best categories and 0.0063 on combination of all categories.

\begin{table*}
\centering
\caption{List of feature categories and their evaluation with XGBoost.}
\label{results}
\begin{tabular}{c|c|cc|cc} 
\hline
& & \multicolumn{2}{c}{Train} & \multicolumn{2}{c}{5-CV} \\
Feature Category & \# Features & Accuracy & Logloss & Accuracy & Logloss \\
\hline
\multicolumn{6}{c}{Hex dump file} \\
\hline
ENT & 203 & 0.9987 & 0.0155 & 0.9862  &  0.0505 \\
1G & 256 & 0.9948 & 0.0307 & 0.9808  &  0.0764 \\
STR & 116 & 0.9877 & 0.0589 & 0.9735  &  0.0993 \\
IMG1 & 52 & 0.9718 & 0.1098 & 0.9550  &  0.1645 \\
IMG2 & 108 & 0.9736 & 0.1230 & 0.9510  & 0.1819  \\
MD1 & 2 & 0.8547 & 0.4043 & 0.8525  & 0.4279  \\
\hline
\multicolumn{6}{c}{disassembled file} \\
\hline
MISC & 95 & 0.9984 & 0.0095 & 0.9917 & 0.0306 \\
OPC & 93 & 0.9973 & 0.0146 & 0.9907  &  0.0405 \\
SEC & 25 & 0.9948 & 0.0217 & 0.9899  &  0.0420 \\
REG & 26 & 0.9932 & 0.0352 & 0.9833  & 0.0695  \\
DP & 24 & 0.9905 & 0.0391 & 0.9811  & 0.0740 \\
API & 796 & 0.9905 & 0.0400 & 0.9843  & 0.0610  \\
SYM & 8 & 0.9815 & 0.0947 & 0.9684 & 0.1372 \\
MD2 & 2 & 0.7655 & 0.6290 & 0.75616  & 0.6621  \\
\hline
\end{tabular}
\end{table*}

\begin{table*}
\centering
\caption{Gradual addition of feature categories based on feature fusion.}
\label{fresults}
\begin{tabular}{c|c|cc|cc}
\hline
& & \multicolumn{2}{c}{Train}  & \multicolumn{2}{c}{5-CV} \\
Feature Category & \# Features & Accuracy & Logloss & Accuracy &  Logloss \\
\hline
C1: MISC+ENT & 298 & 1.0 & 0.0037 & 0.9907 & 0.0322 \\
C2: C1+SEC & 323 & 1.0 & 0.0019 & 0.9920 & 0.0278 \\
C3: C2+API & 1117  & 1.0 & 0.0016 & 0.9927 & 0.0251 \\
C4: C3+1G & 1373  & 1.0 & 0.0015 & 0.9930 & 0.0237 \\
C5: C4+REG & 1399  & 1.0 & 0.0014 & 0.9933 & 0.0226 \\
C6: C5+OPC & 1492 & 1.0 & 0.00137 & 0.9935 & 0.0220 \\
C7: C6+MD1 & 1494 & 1.0 & 0.00132 & 0.9937 & 0.0214 \\
C8: C7+DP & 1518 & 1.0 & 0.00130 & 0.9938 & 0.0210 \\
C9: C8+STR  & 1634  & 1.0 & 0.00128 & 0.9939 & 0.0206 \\
C10: C9+IMG1  & 1686  & 1.0 & 0.00128 & 0.9940 & 0.0203 \\
C11: C10+MD2 & 1688 & 1.0 & 0.00128 & 0.99411 & 0.0201 \\
C12: C11+SYM & 1696 & 1.0 & 0.00128 & 0.99418 & 0.0199 \\
C13: C12+IMG2 & 1804 & 1.0 & 0.00130 & 0.9942 & 0.0197 \\
\hline
\end{tabular}
\end{table*}

\begin{table*}
\centering
\caption{Employing bagging and parameter optimization for XGBoost.}
\label{finalresults}
\begin{tabular}{c|c|cc|c}
\hline
& & \multicolumn{2}{c}{5-CV}  & Test \\
Feature Category & \# Features & Accuracy & Logloss & Logloss \\
\hline
Combination of all categories (C13) & 1804 & \textbf{0.9977} & \textbf{0.0096} & \textbf{0.0063} \\
Combination of best categories (C12) & 1696 & \textbf{0.9976} & \textbf{0.0094} & \textbf{0.0064} \\
\hline
\end{tabular}
\end{table*}


\subsection{Feature extraction time}
\begin{figure}[htp]
\centering
\includegraphics[width=.35\textwidth]{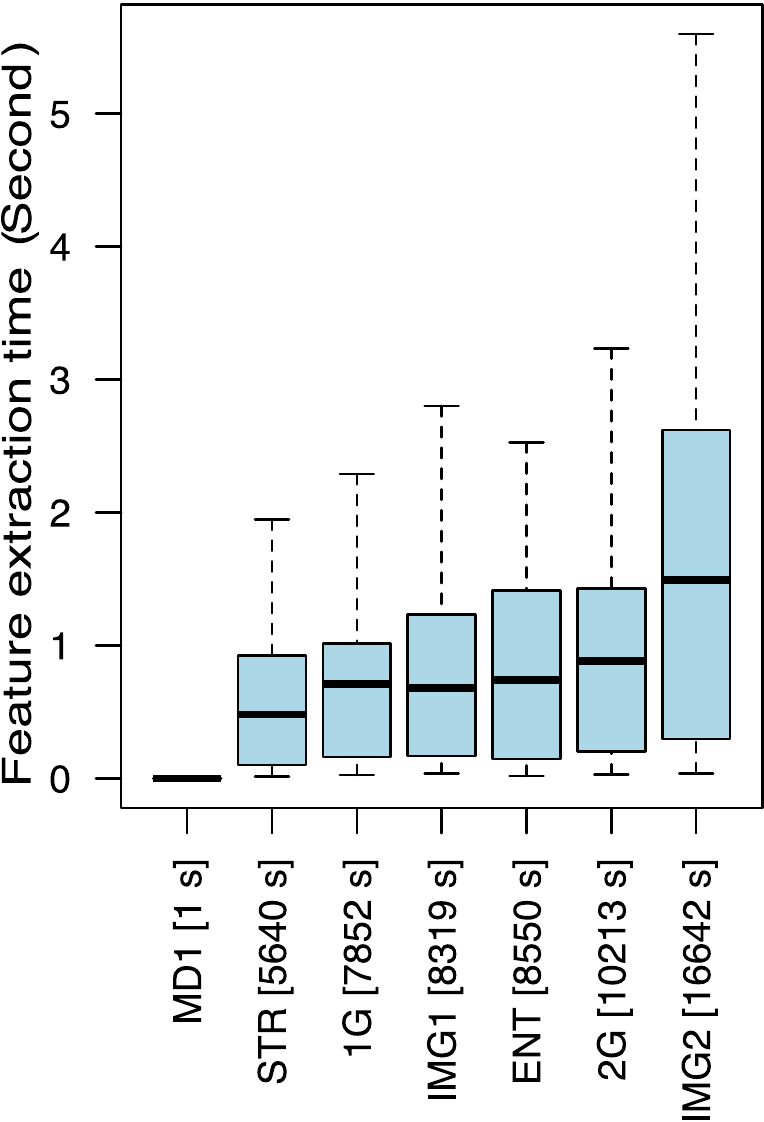}
\caption{The required time of feature extraction from byte code for each app. The time in bracket shows the total time of extraction for all training samples.}
\label{fig:bctime}
\end{figure}

\begin{figure}[htp]
\centering
\includegraphics[width=.35\textwidth]{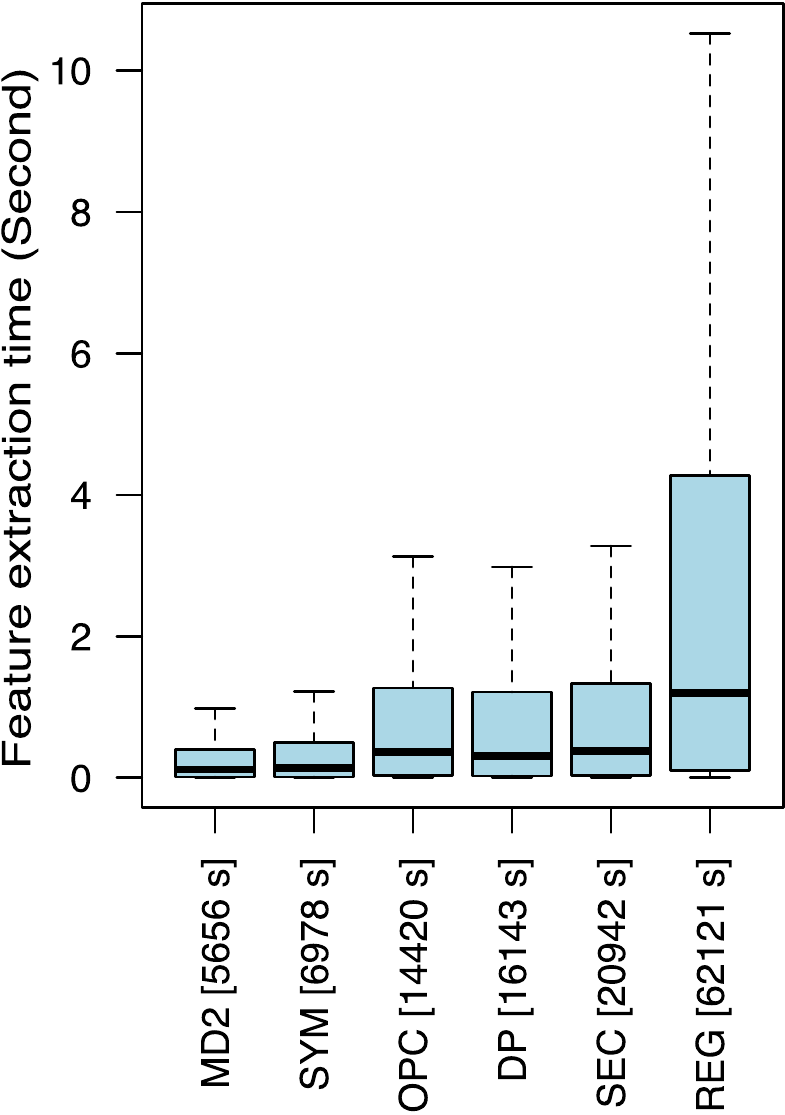}
\caption{The required time of feature extraction from disassembled code for each app. The time in bracket shows the total time of extraction for all training samples.}
\label{fig:asmtime}
\end{figure}

We run the experiments on a laptop with a quad-core processor (2 GHz), and 8GB RAM. Figure~\ref{fig:bctime} and Figure~\ref{fig:asmtime} represents the required time for extracting different feature categories. The tasks of feature extraction and classifier training can be time consuming when the structure of the features is complex, and the size of datasets is large. For example, the 2-Gram category has more than 65K features, which requires a significant amount of time for their extraction (10213 seconds in total in our experiments), for training a model, and for selecting the most relevant ones. As 3-Gram and 4-Gram features are made up of a larger number of components, the time frame required to extract those features is excessively large. 

\subsection{Comparison and Discussion}
To the best of our knowledge, this is the first paper based on the malware dataset that was recently released by Microsoft. Consequently, the effectiveness of the proposed approach can be assessed by comparing the reported results with the ones attained by the winner of the Microsoft malware challenge \footnote{http://blog.kaggle.com/2015/05/26/microsoft-malware-winners-interview-1st-place-no-to-overfitting/}. The winner of the competition attained  0.9983 accuracy, and 0.0031 logloss, on 4-fold cross validation, and 0.0028 logloss on test data, thus confirming the effectiveness of the proposed method, as the significance of this small difference is statistically negligible.
While the performances are quite close, it is worth pointing out the differences between the method proposed in this paper and the one followed by the winning team. The proposed method is characterized by a limited computational complexity compared to the winning method, both in terms of the number and type of features, and in the classification technique employed. 
Firstly, the winning team relied on a large set of well-known features, while we designed the proposed system not only by focusing on the features in the literature that proved to be effective, but also designing novel structural features that could provide a gain in performance with a limited computational cost. As an example, the winning team relied on the extraction of byte code N-gram and operation code N-gram, that require large computational resources both during the training phase, and the testing phase. The complexity of the classification step employed in the proposed method is lower than the ones of the winning team. Both methods rely on the ensemble paradigm, where the winning team resorted to an ensemble of different classifiers in a semi-supervised setting, while we resorted to a standard implementation of XGBoost with bagging. Thus, we can conclude that the proposed method exhibits a better trade-off between computational complexity and performances.

The proposed method has not yet been tested for robustness against evasion attacks \cite{Biggio:2013:test,Kruegel:2005:AMA} or poisoning attacks \cite{Biggio:2014:PBM, biggio:2012:icml} because these kinds of attacks are more frequent against malware detectors rather than against malware classifiers. Attacks against malware classifiers may be used to mislead automatic signature extractors, that analyze malware samples belonging to a family to design effective signatures. As the effectiveness of such attacks depends on a deep knowledge of the malware classifier, as well as of the signature extractor, and this knowledge cannot be reliably inferred from the outside of the system without insider support, we can conclude that these kind of attacks are highly rare. On the other hand, an analysis of the robustness of the system against evasion and poisoning attacks is worth to be carried out if the proposed system is modified to act as a malware detector.

\section{Conclusion and Future work}
We presented a malware classification system characterized by a limited complexity both in feature design and in the classification mechanism employed. To attain this goal, we proposed a number of novel features to represent in a compact way some discriminant characteristics between different families. In particular, we focused on the extraction of novel structural features, that, if compared to content-based features, are easier to compute, and allow the classification of obfuscated and packed malware without the need of deobfuscation and unpacking processes. Reported results allowed assessing the effectiveness of these features both with respect to classification accuracy, and to impurity.

The main motivation behind the choice of a \emph{light} system is its suitability for an industrial use, where the trade-off between complexity and performances can be a key issue. Very often, the gain in performances of complex systems on validation data is negligible compared to the performances of less complex ones. In addition, the use of a reduced set of features may ease the task for an analyst to understand the classification results from the set of features related to a given sample, as compared to complex systems. While we haven't addressed this issue in this paper, we believe in its noteworthiness to gather information about the core common characteristics of malware samples within a family.


%
\bibliographystyle{abbrv}
\bibliography{MicMalChal}  
%
%

\end{document}